\documentclass[aps,prd,reprint,amsmath,amssymb,showpacs,preprintnumbers,superscriptaddress,nofootinbib,longbibliography]{revtex4-1}

\usepackage{slashed}
\usepackage{bm}
\usepackage{latexsym,amssymb,amsmath,float,url,mathrsfs}
\usepackage{bbold}
\usepackage{latexsym}
\usepackage{graphicx}
\usepackage{epstopdf}
\usepackage{amsmath}
\usepackage{subfigure}
\usepackage{hyperref}
\usepackage{pifont}
\usepackage{multirow}
\usepackage[dvipsnames]{xcolor}
\usepackage{tabularx}

\hypersetup{
     colorlinks   = true,
     citecolor    = violet,
     urlcolor     = violet,
     linkcolor    = violet
}

\begin{document}

\title{Enhancing Dark Matter Annihilation Rates with Dark Bremsstrahlung}

\author{Nicole F.\ Bell}
\email{n.bell@unimelb.edu.au}
\affiliation{ARC Centre of Excellence for Particle Physics at the Terascale \\
School of Physics, The University of Melbourne, Victoria 3010, Australia}

\author{Yi Cai}
\email{yi.cai@unimelb.edu.au}
\affiliation{ARC Centre of Excellence for Particle Physics at the Terascale \\
School of Physics, The University of Melbourne, Victoria 3010, Australia}

\author{James B.\ Dent}
\email{jbdent@louisiana.edu}
\affiliation{Department of Physics, University of Louisiana at Lafayette, Lafayette, LA 70504-4210, USA}

\author{Rebecca K.\ Leane}
\email{rebecca.leane@gmail.com}
\affiliation{ARC Centre of Excellence for Particle Physics at the Terascale \\
School of Physics, The University of Melbourne, Victoria 3010, Australia}

\author{Thomas J.\ Weiler}
\email{tom.weiler@vanderbilt.edu}
\affiliation{Department of Physics and Astronomy, Vanderbilt University, Nashville, TN 37235, USA}

\date{\today}

\begin{abstract}
Many dark matter interaction types lead to annihilation processes
which suffer from $p$-wave suppression or helicity suppression,
rendering them subdominant to unsuppressed $s$-wave processes. We
demonstrate that the natural inclusion of dark initial state radiation
can open an unsuppressed $s$-wave annihilation channel, and thus
provide the dominant dark matter annihilation process for particular
interaction types.  We illustrate this effect with the bremsstrahlung
of a dark spin-0 or dark spin-1 particle from fermionic dark matter,
$\overline{\chi}\chi\rightarrow \overline{f}f\phi$ or
$\overline{f}fZ'$.  The dark initial state radiation process, despite
having a 3-body final state, proceeds at the same order in the new
physics scale $\Lambda$ as the annihilation to the 2-body final state
$\overline{\chi}\chi\rightarrow \overline{f}f$.  This opens an unsuppressed $s$-wave at lower order
in $\Lambda$ than the well-studied lifting of helicity suppression via
Standard Model final state radiation, or virtual internal
bremsstrahlung. This dark bremsstrahlung process should influence LHC
and indirect detection searches for dark matter.
\end{abstract}

\maketitle

\section{Introduction}
\label{sec:intro}

The particle nature of dark matter (DM) remains unknown. In order to
significantly probe its properties in indirect detection experiments,
large or unsuppressed annihilation rates are desirable.  The DM
annihilation rate will generally be largest if it proceeds via an
unsuppressed $s$-wave process. Unfortunately, there are a number of
well-motivated DM models in which the $s$-wave annihilation to
Standard Model (SM) products, $\overline{f}f$, is absent or helicity
suppressed.  This renders indirect detection very unlikely, as the
$p$-wave term is suppressed by a factor of the DM velocity squared
(with $v^2 \sim 10^{-6}$ in the present universe) while a helicity
suppression factor of $(m_f/m_\chi)^2$ can be significant for
annihilation to light fermions.  These suppressions are well-known
features of neutralino annihilation in supersymmetric theories, but in
fact are more general.

It is well known that such suppressions can be lifted via the
bremsstrahlung of a SM particle.  For example, an unsuppressed
$s$-wave can be opened via the radiation of a
photon~\cite{Bergstrom:1989jr,Flores:1989ru,Baltz:2002we,Bringmann:2007nk,Bergstrom:2008gr,Barger:2009xe}
or electroweak gauge
boson~\cite{Bell:2010ei,Bell:2011if,Bell:2011eu,Ciafaloni:2011sa,Garny:2011cj,Bell:2012dk,DeSimone:2013gj,Ciafaloni:2013hya}
during the DM annihilation processes.  This has led to much
recent work on the importance of SM radiative corrections in dark
matter
annihilation~\cite{Bergstrom:1989jr,Flores:1989ru,Baltz:2002we,Bringmann:2007nk,Bergstrom:2008gr,Barger:2009xe,Bell:2010ei,Bell:2011if,Bell:2011eu,Ciafaloni:2011sa,Garny:2011cj,Bell:2012dk,DeSimone:2013gj,Ciafaloni:2013hya,Berezinsky:2002hq,Kachelriess:2009zy,Kachelriess:2007aj,Bell:2008ey,Dent:2008qy,Ciafaloni:2010qr,Ciafaloni:2010ti,Barger:2011jg,Fukushima:2012sp,Giacchino:2013bta,Toma:2013bka,Garny:2013ama,Bringmann:2013oja,Bringmann:2015cpa,Bambhaniya:2016cpr,Bambhaniya:2016cpr,Kumar:2016cum,Kumar:2016mrq,Luo:2013bua,Chen:1998dp}.
Despite the bremsstrahlung annihilation process having a 3-body final
state, it can be the dominant annihilation channel in the universe
today (if not at freeze out) because the suppression from additional
coupling and phase space factors is small compared to the $v^2\sim
10^{-6}$ suppression of the $p$-wave contributions. Past work has
primarily used final state radiation (FSR) or virtual internal
bremsstrahlung (VIB) to lift the suppression. If the DM is a SM gauge
singlet, initial state radiation (ISR) of a SM particle is obviously
not possible, however ISR of a $W$ or $Z$ boson from $SU(2)$ charged
DM is possible, and has been considered
in~\cite{Ciafaloni:2011gv,Garny:2011ii,Ciafaloni:2012gs}.

An interesting possibility is that helicity or $p$-wave suppressions
can instead be lifted by the ISR of a {\mbox{\it dark sector}}
field. In this scenario, an initial state dark bremsstrahlung process
can dominate over other suppressed channels. This will require that
the dark sector contains more particles than just the DM candidate
itself which, in fact, is very well motivated: 
the visible sector itself comprises more than one particle species, 
and likewise multiple dark sector fields are a common
feature of many self-consistent, gauge-invariant, and renormalizable
models.  For example, mass generation in the dark sector can require
the introduction of new fields, such as a dark Higgs, while DM
stability may arise from a charge under a new dark sector gauge group,
requiring the introduction of dark photons. More generally, models in
which DM interactions are mediated by the exchange of only an
axial-vector mediator are not gauge invariant. They require the
addition of a dark Higgs to unitarize the longitudinal component of
the gauge boson, and to give mass to both the gauge boson and
DM~\cite{Cline:2014dwa,Kahlhoefer:2015bea,Bell:2016fqf,Duerr:2016tmh,Bell:2016uhg,Duerr:2017uap}. Indeed, the simultaneous presence of both spin-1 and spin-0 mediators lead to
new indirect detection phenomenology that does not arise in single
mediator models \cite{Bell:2016fqf,Duerr:2016tmh,Bell:2016uhg}. Similarly,
both scalar and pseudoscalar mediators can naturally appear together in complete theories~\cite{Goncalves:2016iyg, Bell:2016ekl, Bauer:2017ota, Baek:2017vzd}.

\begin{figure}[h]
\centering
\subfigure{\includegraphics[width=0.45\columnwidth]{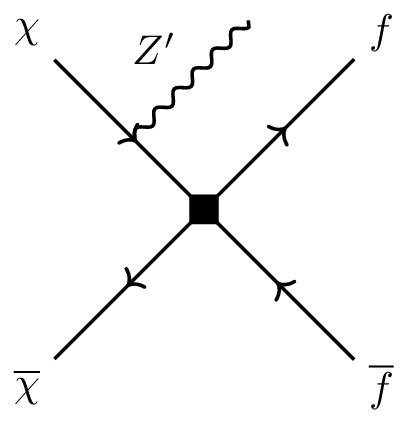}}
\hspace{1mm}
\subfigure{\includegraphics[width=0.45\columnwidth]{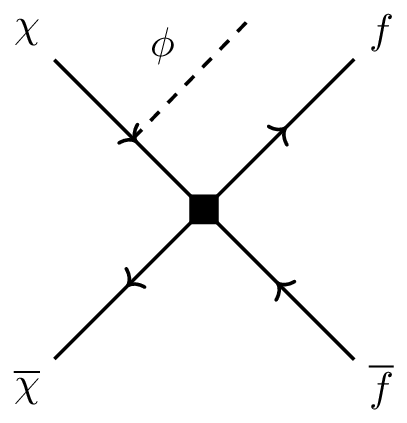}}
\caption{Dark vector (left) and dark scalar (right) ISR. Note in both cases that there is also an additional diagram for emission from the other initial state DM.}
\label{fig:radiation}
\end{figure}

In this paper, for the first time, we explore the possibility that
helicity or $p$-wave suppressions of the DM annihilation process are
lifted by dark bremsstrahlung from the initial state.  We investigate
the case where fermionic DM, $\chi$, radiates either a dark spin-1
field, $Z'$, or spin-0 field, $\phi$, to give the ISR processes
$\overline{\chi}\chi \rightarrow \overline{f}fZ'$ or
$\overline{\chi}\chi \rightarrow \overline{f}f\phi$, respectively, as
shown in Fig.~\ref{fig:radiation}.

Bremsstrahlung annihilation processes are very closely
related to the mono-$X$ processes utilized in collider DM
searches~\cite{ Aaboud:2016tnv,Aaboud:2016uro,Aaboud:2016qgg,ATLAS:2016tsc,ATLAS:1437995,CMS:2016flr,CMS:2016fnh,CMS:2016hmx,CMS:2016pod,CMS:2016mjh,CMS:2016xok,CMS:2016mxc,CMS:2016uxr, Carpenter:2012rg, Bell:2012rg, Carpenter:2013xra, Petrov:2013nia, Berlin:2014cfa, Bell:2015rdw, Bell:2015sza, Haisch:2016usn, Birkedal:2004xn, Gershtein:2008bf, Goodman:2010ku, Goodman:2010yf, Crivellin:2015wva, Petriello:2008pu, Berlin:2014cfa, Lin:2013sca, Fox:2011pm, No:2015xqa, Ghorbani:2016edw, Brennan:2016xjh, Liew:2016oon}, as they are controlled by the same matrix
element. For example, the radiation of photons from fermions in the
FSR annihilation process $\overline{\chi}\chi \rightarrow
\overline{f}f\gamma$ is the analogue of the collider ISR mono-photon
process $\overline{f}f \rightarrow \overline\chi\chi\gamma$.
Likewise, the ISR of a dark spin-0 or spin-1 field from the
initial state $\chi$ in the $\overline\chi\chi  \rightarrow
\overline{f}f\phi$ or $\overline\chi\chi  \rightarrow \overline{f}f Z'$
annihilation processes are then the analogue of the FSR mono-$Z'$
\cite{Bai:2015nfa,Gupta:2015lfa,Autran:2015mfa,Buschmann:2015awa, Zhang:2016sll} or
mono-dark Higgs~\cite{Duerr:2017uap} collider processes, respectively.

For the purpose of illustration, we shall assume the
$\overline\chi\chi\rightarrow \overline{f}f$ process is adequately described by
an effective field theory (EFT operator) of the form $(1/\Lambda^2)
(\overline{\chi}\Gamma\chi)(\overline{f}\Gamma f)$.  We will see that
the $s$-wave contribution to the ISR process scales as $\langle\sigma
v\rangle_{\rm ISR} \propto \mathcal{O}(1/\Lambda^4)$, i.e., the same
order in $\Lambda$ as the 2-body annihilation $\overline\chi\chi \rightarrow
\overline{f}f$.  In comparison, the well-studied lifting of helicity
suppressions via FSR or VIB radiation can only produce unsuppressed $s$-wave cross sections
at higher order in $1/\Lambda$, with cross sections scaling as $\langle\sigma
v\rangle_{\rm FSR, \,ISR} \propto
\mathcal{O}(1/\Lambda^8)$\footnote{This observation was also made in
  the case of ISR of a $W/Z$ boson from $SU(2)$ doublet
  DM~\cite{Ciafaloni:2011gv,Garny:2011ii,Ciafaloni:2012gs}.}.

In Section \ref{sec:fermionicann}, we provide an overview of
suppressions to fermionic DM annihilation cross sections, and
discuss annihilation both directly to SM particles, and to dark
mediators. In Section \ref{sec:isr} we outline possible dark ISR
annihilation processes, and investigate two interesting cases in more
detail in Sections \ref{sec:aavec} and \ref{sec:pseudoscalar}. We
present our conclusions in Section \ref{sec:conclusion}.

\section{Overview of Fermionic Dark Matter Annihilation}
\label{sec:fermionicann}

\subsection{Direct annihilation to SM particles}

If DM is a Majorana fermion, the possible interactions which can
mediate a $\chi\chi\rightarrow \overline{f}f$ annihilation process
are:
\begin{itemize}
\item
{$s$-channel exchange of an axial-vector}: helicity suppressed $s$-wave,
\item
{$s$-channel exchange of a scalar}: no $s$-wave,
\item
{$s$-channel exchange of a pseudoscalar}: unsuppressed $s$-wave, or
\item
{$t$-channel exchange of a sfermion-like scalar}: helicity suppressed $s$-wave.
\end{itemize}
In the $t$-channel case, Fierz rearrangement to $s$-channel form gives
$A\otimes A$ and $A\otimes V$ structures.  The $A\otimes A$ has a
helicity suppressed $s$-wave, while the $A\otimes V$ has no
$s$-wave. For Majorana DM, we thus see that the $s$-channel exchange
of a pseudoscalar is the only case of an unsuppressed $s$-wave.  All
other possibilities feature either helicity or $v^2$ suppressions.

For Dirac DM, there are additional possibilities because vector
couplings, which vanish for Majorana particles, are also allowed.  Note,
however, that while the exchange of a vector results in an
unsuppressed $s$-wave annihilation cross section, these models are also
well constrained because they lead to unsuppressed spin-independent scattering in
direct detection experiments.

A summary of the cross section suppression factors, for both annihilation and scattering, for all possible Lorentz structures for $\overline\chi\chi\rightarrow \overline{f}f$, is given in Ref.~\cite{Kumar:2013iva}.

\subsection{Direct annihilation to dark mediators}

Table \ref{tab:tchannelnew} details whether fermionic DM annihilation to two different mediators ($\overline\chi\chi\rightarrow M_1 M_2$) is $s$- or $p$-wave, depending on the Lorentz structures of the DM-mediator interactions. For annihilation to any two spin-1 mediators, the rate is $s$-wave. For any two spin-0 mediators, the rate is $p$-wave unless one scalar and one pseudoscalar are both present. For a mixed spin-0 and spin-1 final state, if the spin-1 is a vector, the processes are $s$-wave, while if the spin-1 is an axial-vector, the processes are $p$-wave\footnote{If there exist direct couplings between the dark mediators themselves, it is possible to have an $s$-wave process with a spin-0 plus spin-1 final state where the spin-1 is an axial-vector~\cite{Bell:2016fqf, Bell:2016uhg}. Details of such couplings are model dependent, and we do not consider such processes in this work.}.

If one of these mediators is off-shell while the other is on-shell, it
is equivalent to the dark ISR process discussed in the following
section --- where the on-shell mediator corresponds to the dark ISR, and the off-shell mediator has been integrated out to give
the EFT vertex. As such, the annihilation type for dark ISR is related
to the underlying Lorentz structures of the mediators. We now discuss
the dark ISR processes in detail.

\begin{table*}
\centering
\def\arraystretch{1.6}
\resizebox{0.88\textwidth}{!}{
\begin{tabular}{|c||c|c|c||c|c|c||c|c|c|c|}
\hline
$\Gamma_{M_1} \otimes \Gamma_{M_2}$ & $S\otimes S$ & $S\otimes P$ & $P\otimes P$ & $V\otimes V$ & $V\otimes A$ & $A\otimes A$ & $S\otimes V$ & $S\otimes A$ & $P\otimes V$ & $P\otimes A$ \\
\hline
\hline
$\overline{\chi}\chi\rightarrow M_1 \, M_2$ &  $v^2$ & 1 & $v^2$ & 1 & 1 & 1 & 1 & $v^2$ & 1 & $v^2$ \\
\hline
\end{tabular}}
\caption{Suppression factors for fermionic DM annihilation to two different mediators $M_1$ and $M_2$, which have varying Lorentz structures: vector (V), axial-vector (A), scalar (S) or pseudoscalar (P). The combination of mediators can be two spin-0 final states, two spin-1 final states, or a mixed spin-0 plus spin-1 final state.  Note for Majorana DM, the $V$ cases do not exist.}
\label{tab:tchannelnew}
\end{table*}
\begin{table*}
\centering
\def\arraystretch{1.6}
\resizebox{0.6\textwidth}{!}{
\begin{tabular}{|c|c|c|c|c|c|}
\hline
\multirow{2}{*}{$\Gamma_\chi\otimes\Gamma_f$} & \multirow{2}{*}{$\overline{\chi}\chi\rightarrow \overline{f}f$} & \multicolumn{2}{ c| } {$\overline{\chi}\chi\rightarrow \overline{f}fZ'$} & \multicolumn{2}{ c| }{$\overline{\chi}\chi\rightarrow \overline{f}f\phi$} \\
\cline{3-6}
 &  & $\Gamma_{Z'}=V$ & $\Gamma_{Z'}=A$ & $\Gamma_{\phi}=S$ & $\Gamma_{\phi}=P$ \\ \hline \hline
$V\otimes V$ & 1 & 1 & 1 & 1 & 1 \\ \hline
$A\otimes V$ & $v^2$ & 1 & 1 & $v^2$ & $v^2$  \\ \hline
$V\otimes A$ & 1 & 1 & 1 & 1 & 1 \\ \hline
$A\otimes A$ & $(m_f/m_\chi)^2$ & 1 & 1 & $v^2$ & $v^2$  \\ \hline
$S\otimes S$ & $v^2$ & 1 & $v^2$ & $v^2$ & 1 \\ \hline
$P\otimes S$ & 1 & 1 & $v^2$ & $1^*$ & $v^2$  \\ \hline
$S\otimes P$ & $v^2$ & 1 & $v^2$ & $v^2$ & $1^*$   \\ \hline
$P\otimes P$ & 1 & 1 & $v^2$ & 1 & $v^2$ \\ \hline
\end{tabular}}
\caption{Suppression factors for fermionic DM annihilation processes,
  for varying Lorentz structures: vector (V), axial-vector (A), scalar
  (S) or pseudoscalar (P). In some cases, helicity suppression
  ($\propto m_f^2/m_\chi^2 $) or $p$-wave suppression ($\propto v^2$)
  can be lifted to $s$-wave ($\propto 1$) by including dark
  bremsstrahlung, with no relative suppression factors.  Note that the
  presence of any vector or axial-vector ISR implies the existence of
  the competing $s$-wave process $\overline\chi\chi\rightarrow Z'Z'$.
  An asterisk $*$ indicates that the $s$-wave process
  $\overline\chi\chi\rightarrow \phi_{1}\phi_{2}$ will inevitably be
  induced via scalar($\phi_{1}$)-pseudoscalar($\phi_{2}$) mixing in a
  CP violating scenario.  Note for Majorana DM, the $V$ cases do not
  exist.}
\label{table:ISR}
\end{table*}

\section{Dark Initial State Radiation}
\label{sec:isr}

In this section, we consider the scenario where the ISR of a dark sector particle lifts helicity or $p$-wave suppression in 
fermionic DM annihilation processes.

Figure~\ref{fig:radiation} demonstrates the dark sector ISR in DM annihilation. For the sake of illustration, 
we use an EFT to describe the interactions between DM and SM fermions. The qualitative effects we discuss are relevant for UV completions which map to the relevant cases.
We assume one mediator is sufficiently heavy, such that the EFT description can safely be used without unitarity issues.

Table~\ref{table:ISR} details the annihilation type and relative suppression of all processes (whether they are $s$-wave, $p$-wave, or helicity suppressed). This reveals which Lorentz structures for particular dark ISR will lift suppression in DM annihilation.

We see, for example, that the radiation of a dark vector is a promising ISR scenario which lifts the suppression of DM annihilation for several Lorentz structures: $S\otimes S$, $S\otimes P$, $A\otimes A$, and $A\otimes V$. Radiating an axial-vector lifts suppression in $A\otimes A$ and $A\otimes V$ annihilation processes. Radiating a scalar fails to lift any suppression of the annihilation cross section. Radiating a pseudoscalar, however, makes a process with a $S\otimes S$ or $S\otimes P$ structure $s$-wave. In the case of $S\otimes P$ or $P\otimes S$, any scalar with such a structure will not have well-defined CP properties. Thus the mixing between the heavy scalar and the pseudoscalar is inevitable, and a $2\rightarrow2$ $s$-wave contribution, $\overline\chi\chi\rightarrow \phi_{1}\phi_{2}$ will be induced, where $\phi_{1}$ is a scalar and $\phi_{2}$ is a pseudoscalar.

It is also important to note that once an additional dark sector
field is included to allow dark ISR, there can also be $s$-wave
annihilations of DM into the dark radiation. For spin-1 ISR, the
direct annihilation to mediators $\overline\chi\chi\rightarrow Z'Z'$
is $s$-wave for both vector and axial-vector couplings, and can
dominate the total DM annihilation rate for some choices of the
coupling strength or masses. In the case that the dark radiation is a
\mbox{spin-0} field, the $t$-channel annihilation process
$\overline\chi\chi\rightarrow \phi\phi$ is $p$-wave suppressed for
both scalar and pseudoscalar couplings, and so can very naturally be
subdominant to the suppression-lifting ISR process.
Also note that for sufficiently light dark radiation, Sommerfeld effects can be important.

To avoid the ``dark radiation'' contributing to the relic
density, it must eventually decay to SM states.  This can easily be
arranged without introducing other consequences, e.g., via a gauge or
Higgs portal to the SM, which can naturally appear for inclusion of a
gauge boson or scalar, respectively. Such SM states can be a signal for
indirect detection experiments (see, i.e., Refs.~\cite{Bell:2016fqf, Berlin:2015wwa}). We will assume that the 
couplings of the dark radiation to the SM are small, such that the $2\rightarrow2$ 
exchange of dark radiation will be subdominant.

We now study in detail two particular cases of the lifting helicity or
$p$-wave suppression: $S\otimes S$ with dark pseudoscalar ISR and
$A\otimes A$ with dark vector ISR.  We choose the former as it is the
only scenario where introducing dark ISR to lift a $p$-wave cross
section does not induce an additional competing $2\rightarrow2$
$s$-wave process.  We choose the latter as an example of lifting
helicity suppression.  Other scenarios and Lorentz structures can
dominate in particular regions of parameter space. Note also that in
all the scenarios discussed, UV completions with the same Lorentz
structures would map to the same results we present. Our results are
not specific to EFTs, but rather to the underlying Lorentz structures.

\section{ Lifting P-wave suppression in $S\otimes S$ interactions}
\label{sec:pseudoscalar}

In this section, we demonstrate how $p$-wave suppression can be lifted
through dark pseudoscalar ISR, in the case of the Lorentz structure
$\Gamma_\chi\otimes\Gamma_f=S\otimes S$. Such a structure is possible
for both Majorana and Dirac DM, with the Majorana interaction terms
differing by a factor of 1/2. We will also discuss any new competing
annihilation processes.

\subsection{$p$-wave suppressed $\overline\chi\chi\rightarrow \overline{f}f$}

For the Lorentz structure $\Gamma_\chi\otimes\Gamma_f=S\otimes S$, the DM interactions with SM fermions are described by the four-Fermi operator
\begin{equation}
\mathcal{L}_{\rm int}\supset\frac{1}{\Lambda^2}(\overline{\chi}\chi)(\overline{f}f),
\label{eq:scalar_eft}
\end{equation}
where $\chi$ is a Dirac DM candidate, $f$ are SM fermions and $\Lambda$ is the cutoff scale for new physics, representing a heavy field which has been integrated out.

The operator in Eq.~(\ref{eq:scalar_eft}) yields a $p$-wave suppressed
DM annihilation cross section for $\overline{\chi}\chi\rightarrow
\overline{f}f$ given by
\begin{equation}
 \sigma v = \frac{v^2 m_{\chi }^2 \left(1-m_f^2/m_{\chi}^2\right){}^{3/2}}{8 \pi  \Lambda^4}.
 \label{eq:phiff}
\end{equation}
where $m_f$ is the mass of the SM fermion and $m_\chi$ is the DM mass. 
$v$ here denotes relative velocity. The $v^2$ prefactor shows that this is clearly a $p$-wave suppressed process. We now explicitly show that, for such an operator, including
dark pseudoscalar radiation lifts this suppression, and the dominant
$s$-wave process can be $\overline{\chi}\chi\rightarrow
\overline{f}f\phi$.

\subsection{Dark pseudoscalar ISR, $\overline\chi\chi\rightarrow \overline{f}f\phi$}

We consider a minimal setup in which the EFT operator of
Eq.~(\ref{eq:scalar_eft}) is augmented by a coupling of the DM to a new
pseudoscalar $\phi$,
\begin{equation}
 \mathcal{L}_{\rm int}\supset\frac{1}{\Lambda^2}(\overline{\chi}\chi)(\overline{f}f)+i\,g_\phi\phi\overline{\chi}\gamma_5\chi, 
 \label{eq:lag_scalar}
\end{equation}
where $g_\phi$ is the coupling constant.  While some complete model
may have relations between the couplings and masses of the dark sector
particles, for the sake of illustration we take all masses and
couplings to be independent parameters.  The annihilation cross
section for the dark pseudoscalar ISR process
$\overline{\chi}\chi\rightarrow \overline{f}f\phi$ is given (in the
$m_f=0$ limit) by
\begin{align}
&\langle\sigma v\rangle_{\chi\overline{\chi}\rightarrow f\overline{f}\phi} = \frac{g_\phi^2 m_\chi^2}{48 \pi ^3 \Lambda^4}\times\\
&\bigg\{1+24\rho_\phi^3\sqrt{1-\rho_\phi^2}(5\rho_\phi^2-2)\tan^{-1}\frac{\sqrt{1-\rho_\phi^2}}{\rho_\phi}\nonumber\\
&+21 \rho_\phi^2-105\rho_\phi^4 +83\rho_\phi^6 + 12 \rho_\phi^2(1 - 9\rho_\phi^2+10 \rho_\phi^4) \ln\rho_\phi \bigg\},\nonumber
\end{align}
where $\rho_\phi= m_\phi/2 m_\chi$. 
Clearly this process is no longer velocity suppressed term, and so is $s$-wave. Furthermore, it scales $\propto \mathcal{O}(1/\Lambda^4)$, i.e., the same  order in $\Lambda$ as the annihilation to $\overline{f}f$ which scales as $\propto \mathcal{O}(v^2/\Lambda^4)$. In contrast, FSR or VIB of a SM particle, e.g. $\overline{\chi}\chi\rightarrow\overline{f}f\gamma$, only allows unsuppressed $s$-wave annihilation at higher order in $1/\Lambda$, with a cross section scaling $\propto \mathcal{O}(1/\Lambda^8)$.

\subsection{Competition with $\overline\chi\chi\rightarrow \phi\phi\phi$}

\begin{figure}[b]
\includegraphics[width=0.55\columnwidth]{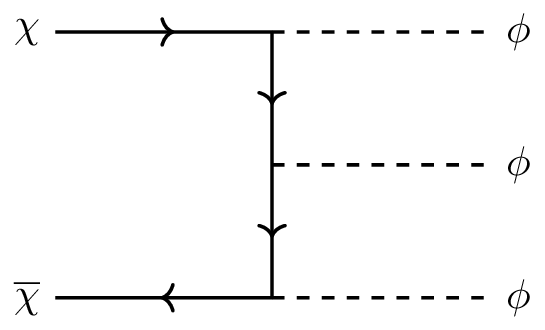}
\caption{$s$-wave process for DM annihilation to pseudoscalars, $\overline\chi\chi\rightarrow \phi\phi\phi$. Note that there are a total of six diagrams that contribute.}
\label{fig:3scalar}
\end{figure}

Unlike the dark vector radiation case which will be discussed in the next
section, there is no additional $s$-wave $2\rightarrow2$ process
induced for spin-0 fields because the $\phi\phi$ final state is $p$-wave
suppressed. In the limit $m_\phi\ll m_\chi$, the $p$-wave suppressed
cross section for $\overline\chi\chi\rightarrow \phi\phi$ is
\begin{equation}
 \sigma v_{\overline\chi\chi\rightarrow\phi\phi} \simeq \frac{g_\phi^4v^2}{384\pi m_{\chi}^2} \; .
 \label{eq:phiphi}
\end{equation}
Instead, the $s$-wave annihilation to 3 pseudoscalars
$\phi\phi\phi$~\cite{Abdullah:2014lla} shown in Fig.~\ref{fig:3scalar}
will compete with the ISR $\overline{f}f\phi$ channel.  Note that both
these $2\rightarrow 3$ process suffer the same 3-body phase space
suppression. In the limit $m_\phi\ll m_\chi$, the $s$-wave cross section for $\overline\chi\chi\rightarrow \phi\phi\phi$ is
\begin{align}
\langle \sigma v\rangle_{\overline\chi\chi\rightarrow \phi\phi\phi} \simeq \frac{g_\phi^6(7\pi^2-60)}{1536 \pi^3 m_\chi^2 } \; .
\label{eq:massless3pseudo}
\end{align}
The full cross section is described in Appendix~\ref{sec:full3pseudo}.

Figure~\ref{fig:scalar_compare} displays the annihilation cross
sections for the pseudoscalar ISR and $\phi\phi\phi$ annihilation
processes, illustrating potential regions of parameter space where
either process is the dominant annihilation channel, depending on
values of the couplings, DM and pseudoscalar masses. Other than
providing kinematic thresholds for when the annihilations are allowed,
the rates are effectively independent of the pseudoscalar mass. The
pseudoscalar ISR process dominates over the three-body pseudoscalar
process for much of the parameter space.

\begin{figure}[t]
\includegraphics[width=\columnwidth]{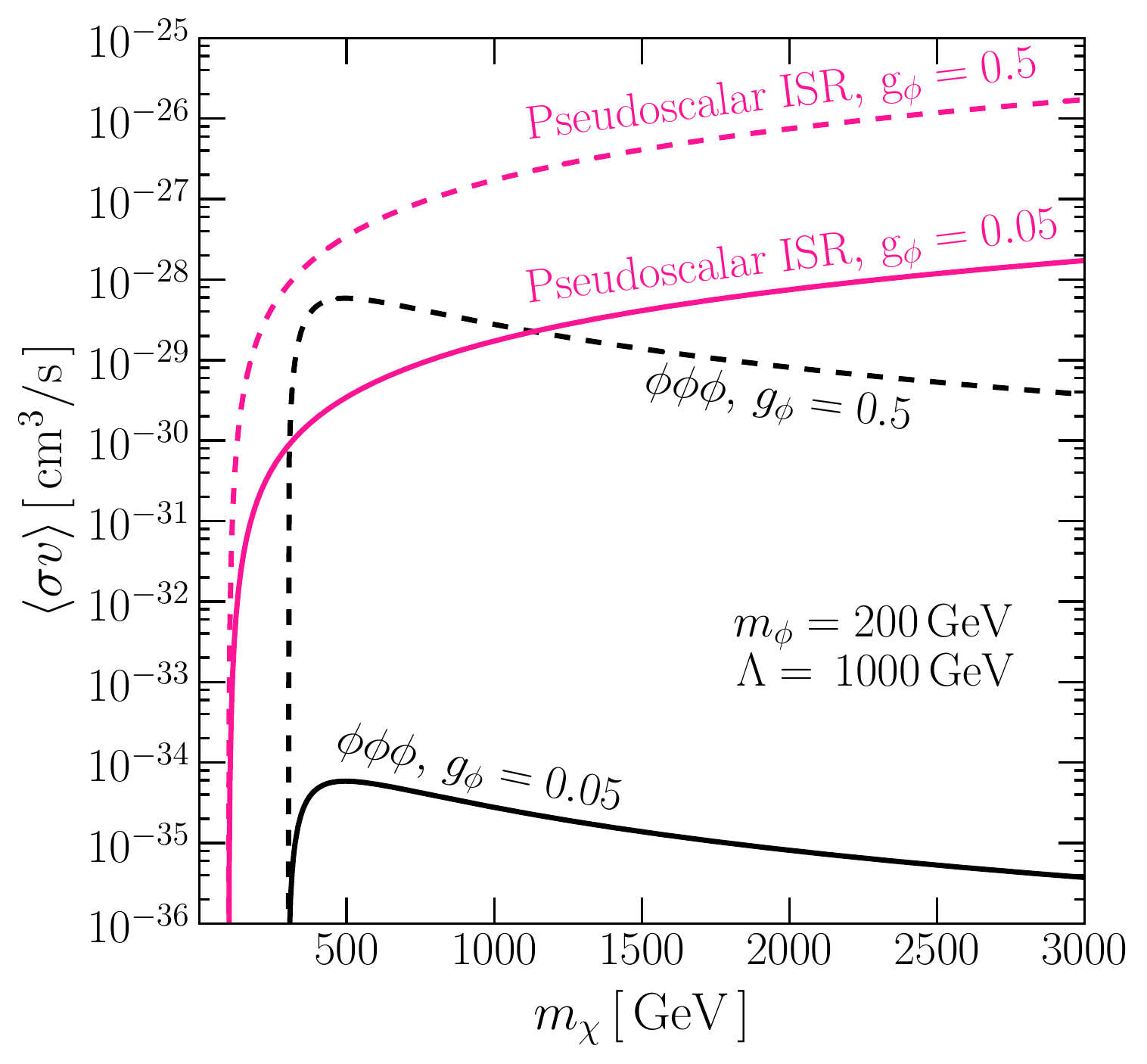}\\
\vspace{0.5cm}
\includegraphics[width=\columnwidth]{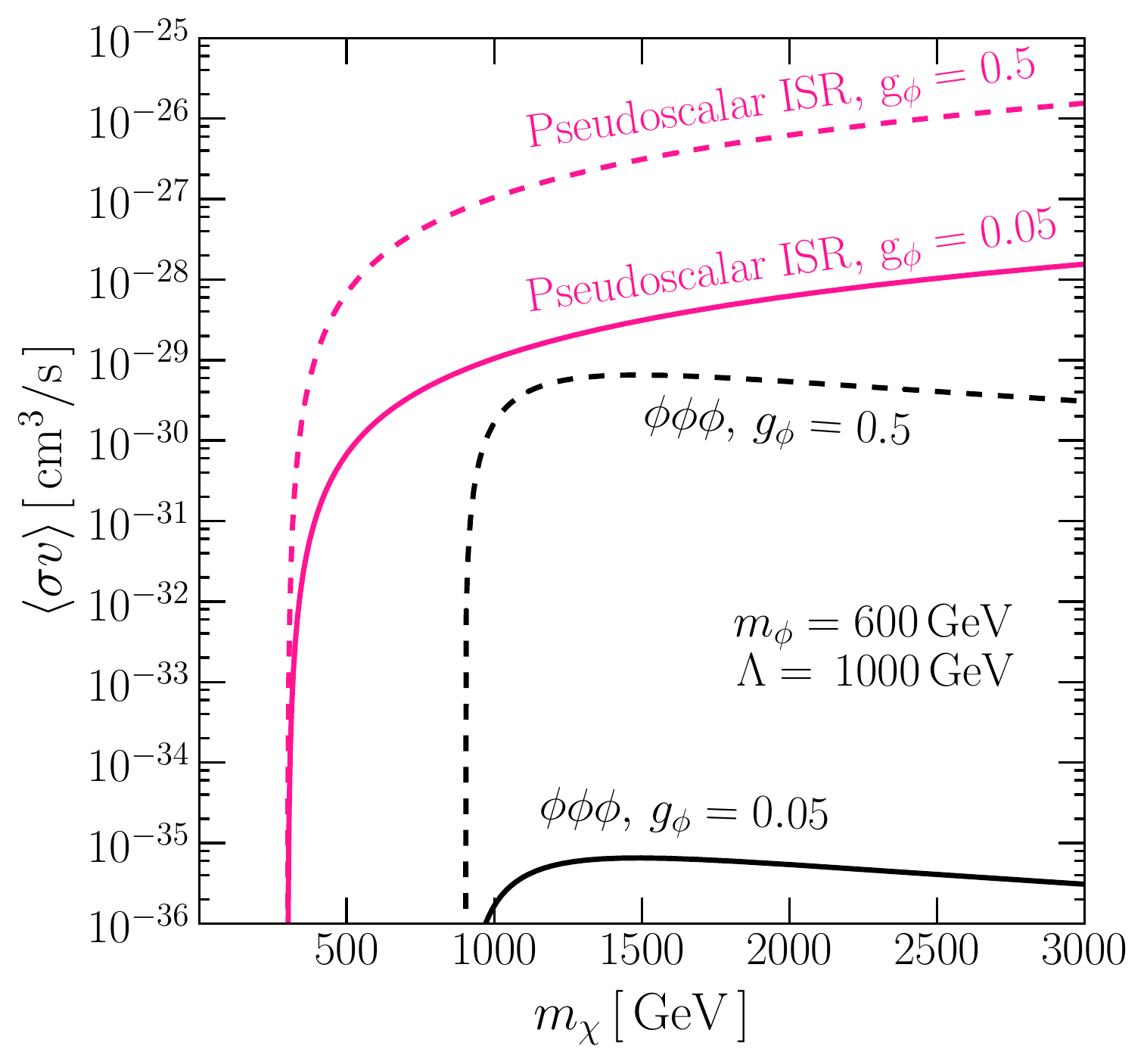}
\caption{Comparison of the $s$-wave cross sections for the
  pseudoscalar ISR (pink) and 3-body $\phi\phi\phi$ (black)
  annihilation processes for $m_{\phi}$, $\Lambda$ and $g_\phi$ as
  labeled, and $m_f=0$. Note the largest value shown for $m_\chi$
  corresponds to $m_\chi \sim \sqrt{4\pi} \Lambda$, where the validity
  of the EFT within the perturbative regime is satisfied.}
\label{fig:scalar_compare}
\end{figure}

\section{Lifting Helicity Suppression in $A\otimes A$ Interactions}
\label{sec:aavec}

In this section, we demonstrate how helicity suppression can be lifted
through dark vector ISR, in the case of the Lorentz structure
$\Gamma_\chi\otimes\Gamma_f=A\otimes A$. Such a structure is very
natural for Majorana DM, but is also possible with Dirac DM. We will
also discuss any new competing annihilation processes.

\subsection{Helicity suppressed $\overline{\chi}\chi\rightarrow \overline{f}f$}

For the Lorentz structure $\Gamma_\chi\otimes\Gamma_f=A\otimes A$, the
DM interactions with SM fermions are described by the four-Fermi
operator
\begin{equation}
\mathcal{L}_{\rm int}\supset\frac{1}{\Lambda^2}(\overline{\chi}\gamma^\mu\gamma^5\chi)(\overline{f}\gamma^\mu\gamma^5 f),
\label{eq:axa_eft}
\end{equation}
where we shall again assume that the DM candidate $\chi$ is a Dirac
fermion.

The operator in Eq.~(\ref{eq:axa_eft}) yields a helicity suppressed DM
annihilation cross section for $\overline{\chi}\chi\rightarrow
\overline{f}f$,
\begin{equation}
 \langle\sigma v\rangle_{\overline{\chi}\chi\rightarrow \overline{f}f}=\frac{m_f^2 \sqrt{1-m_f^2/m_{\chi }^2}}{2 \pi  \Lambda ^4}.
\end{equation}
In the limit that $m_f\rightarrow0$, this process is vanishing. We now
explicitly show that including dark vector radiation of a $Z'$ lifts
this helicity suppression, such that the dominant $s$-wave process can be
$\overline{\chi}\chi\rightarrow \overline{f}fZ'$.

\subsection{Dark vector ISR, $\overline{\chi}\chi\rightarrow \overline{f}fZ'$}

We again consider a minimal scenario where, in addition to the EFT
interaction of Eq.~(\ref{eq:axa_eft}), we include a coupling of the DM
to a new spin-1 field, $Z^\prime$,
\begin{align}
\mathcal{L}_{\rm int} \supset \frac{1}{\Lambda^2} \left(\bar{\chi} \gamma_\mu \gamma_5 \chi \right)\left( \overline{f} \gamma^\mu \gamma_5 f  \right) +  g_{Z'}\overline{\chi} \gamma^\mu \chi Z'_\mu \; , 
   \label{eq:axa_lag}
\end{align}
where $g_{Z'}$ is the coupling constant.  For the sake of
illustration, we have chosen a $Z'$ coupling of vector form to Dirac
DM. Alternatively, an axial-vector $Z'$ could be chosen (for either
Dirac and Majorana DM) which would also open an unsuppressed $s$-wave,
as shown in Table~\ref{table:ISR}.

The annihilation cross section for the dark vector ISR process
$\overline{\chi}\chi\rightarrow \overline{f}fZ'$ is given (in the
$m_f=0$ limit) by
\begin{align}
&\langle\sigma v\rangle_{\overline{\chi}\chi\rightarrow  \overline{f}fZ'} = \frac{g_{Z'}^2 m_\chi^2}{36 \pi^3 \Lambda^4}\times\\
&\bigg\{4+24 \rho_{Z'}^3(1+5\rho_{Z'}^2)\sqrt{1-\rho_{Z'}^2} \tan^{-1} \frac{\sqrt{1-\rho_{Z'}^2}}{\rho_{Z'}}\nonumber\\
&-27\rho_{Z'}^2-60\rho_{Z'}^4+83\rho_{Z'}^6 + 12\rho_{Z'}^4(10\rho_{Z'}^2-3) \ln \rho_{Z'}
\bigg\},\nonumber
\end{align}
where $\rho_{Z'}=m_{Z'}/2m_\chi$.  As the cross section does not vanish
in the limit $m_f\to 0$, this process no longer has the
helicity suppressed $(m_f^2/m_\chi^2)$ dependence. It also has no
velocity suppression, and so is $s$-wave. Again, we see the dark ISR cross
section scales $\propto \mathcal{O}(1/\Lambda^4)$, i.e., the same
order in $\Lambda$ as the annihilation to $\overline{f}f$ which scales
$\propto \mathcal{O}(m_f^2/(m_\chi^2\Lambda^4))$.

\subsection{Competition with $\overline\chi\chi\rightarrow Z'Z'$}
\begin{figure}[H]
\centering
\includegraphics[width=0.45\columnwidth]{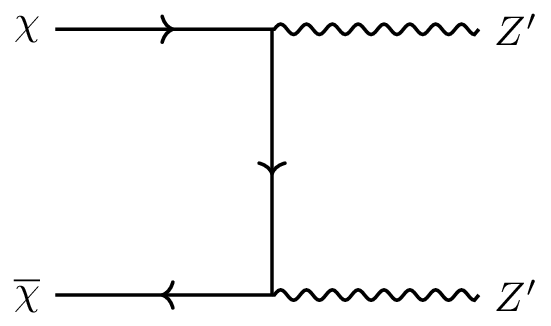}
\caption{$s$-wave process for DM annihilation to dark vectors. Note there is also a contribution from the $u$-channel diagram.}
\label{fig:zpzp}
\end{figure}

The inclusion of the $Z'$ vector induces an additional two-body
annihilation process, $\overline\chi\chi \to Z'Z'$, as shown in Fig.~\ref{fig:zpzp}. This process is
also $s$-wave (irrespective of whether the $Z'$ couplings are of
vector or axial-vector form). Therefore, the dominant annihilation
channel will be $\overline\chi\chi\to Z'Z'$ or $\overline\chi\chi\to
\overline{f}fZ'$, either of which may dominate depending on the region
of parameter space.  The cross section for the annihilation of Dirac
DM to a pair of $Z'$ is given by
\begin{align}
\langle\sigma v\rangle \, _{\overline{\chi}\chi\rightarrow Z'Z'}=\frac{g_{Z'}^4\left(1-4\rho_{Z'}^2\right)^{\frac{3}{2}}}{16\pi m_\chi^2 (1-2\rho_{Z'}^2)^2},
\end{align} 
where again $\rho_{Z'}=m_{Z'}/2m_\chi$.

\begin{figure}[H]
     \begin{center}
       \includegraphics[width=\columnwidth]{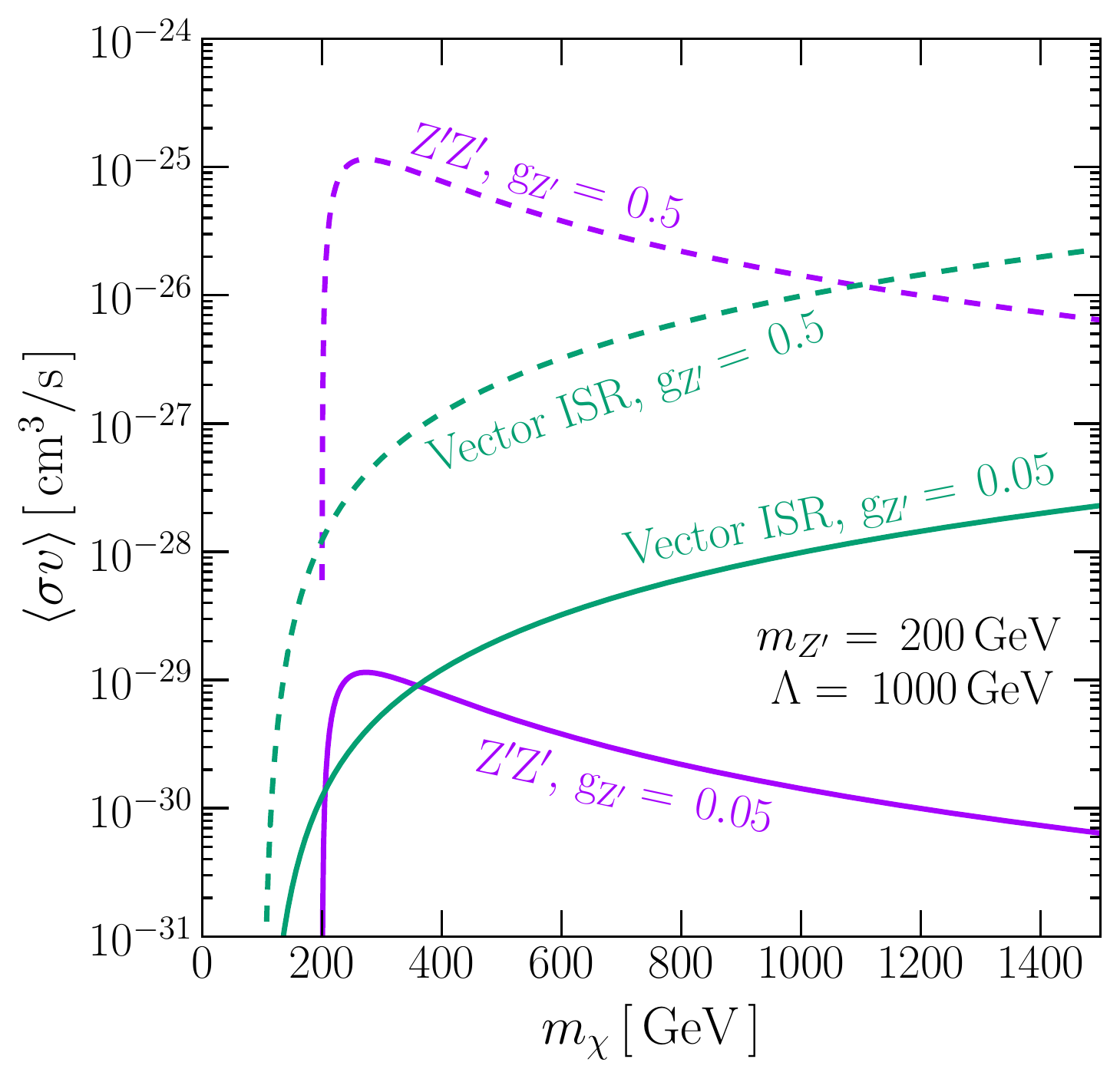}\\
       \vspace{0.5cm}
       \includegraphics[width=\columnwidth]{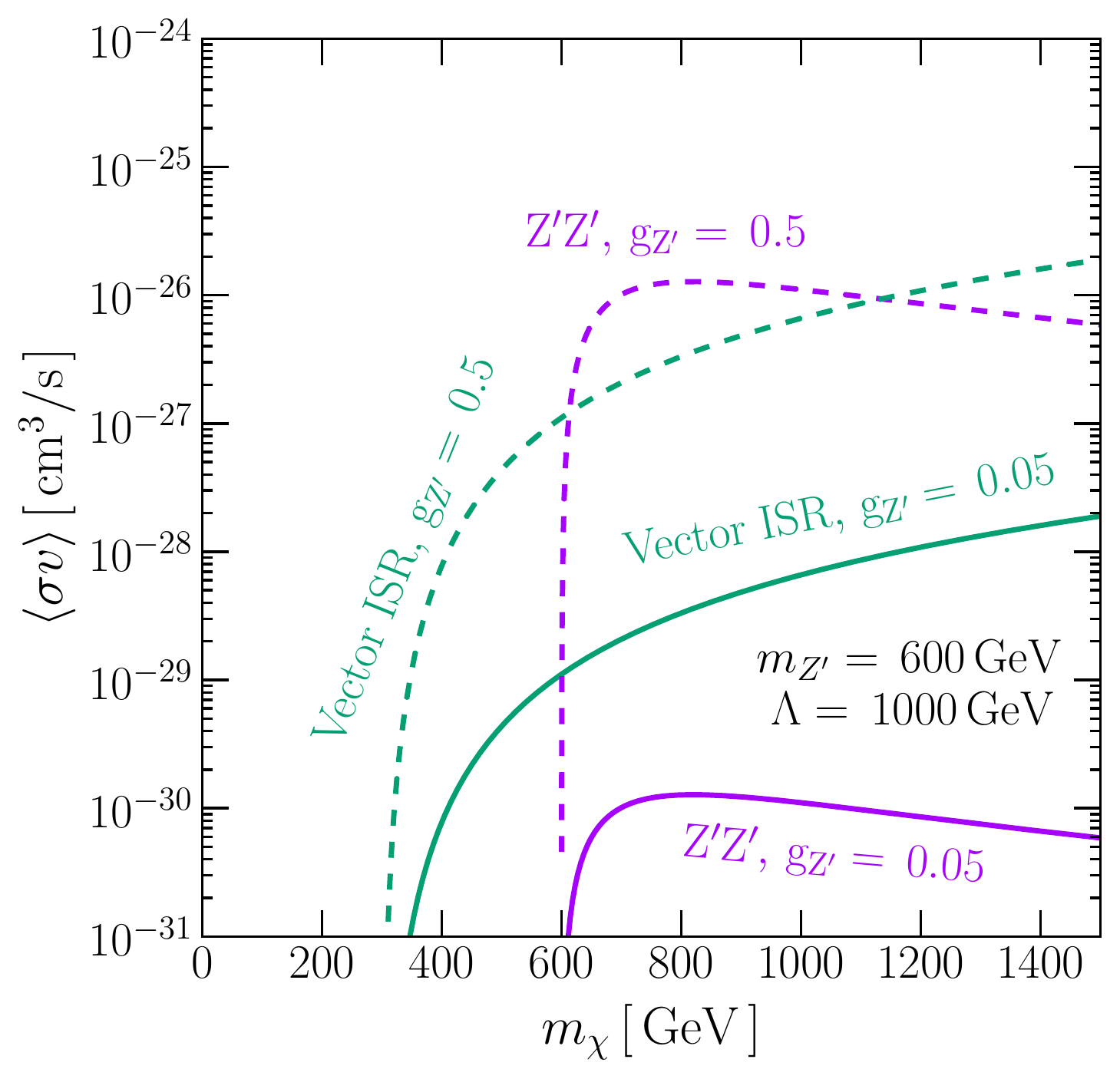}
    \end{center}
\caption{Comparison of the $s$-wave cross sections for Dirac DM annihilating via the dark vector ISR process (cyan) and the
  2-body $Z'Z'$ process (purple), for $m_{Z'}$, $\Lambda$ and $g_{Z'}$
  as labeled, and $m_f=0$. Note the largest value shown for
  $m_\chi$ corresponds approximately to the largest value permitted within a gauge-invariant and perturbative framework, without other new physics appearing.}
\label{fig:axa_compare}
\end{figure}

Figure~\ref{fig:axa_compare} displays the annihilation cross sections
for Dirac DM for both processes, illustrating regions of parameter
space where either process may be the dominant annihilation channel.
As the annihilation cross section to $\overline{f}fZ'$ contains two
powers of $g_{Z'}$ while that for $Z'Z'$ contains fours powers of this
coupling, the dark ISR process can easily dominate when $g_{Z'}$ is small.  As
the coupling becomes larger, $Z'Z'$ dominates for more of the
parameter space. If the $Z'$ is particularly heavy, the dark ISR
process can also dominate over $Z'Z'$ due to kinematic constraints.

\section{Conclusion}
\label{sec:conclusion}

The observation of an unexplained flux of SM particles in the
astrophysical sky can be interpreted as a DM signal. To probe the
nature of DM via such an indirect detection signal, it is important to
know which processes may provide the dominant contributions or
strongest constraints.  In this paper, for the first time, we have
explored the possibility that dark ISR can open an unsuppressed
$s$-wave annihilation channel.  This can be the dominant DM
annihilation mode in models where the lowest order processes are
helicity or $p$-wave suppressed.

We found that dark ISR from the initial state $\overline{\chi}\chi$
can lift such suppressions for several different types of dark
radiation and DM interaction structures.  For four-Fermi type
interactions of DM with SM fermions, several Lorentz structures suffer
suppressed $2\rightarrow 2$ annihilation processes --- $A\otimes A$,
$A\otimes V$, $S\otimes S$, and $S\otimes P$.  The ISR of a dark
vector opens an unsuppressed $s$-wave in all these cases.  Radiating
an axial-vector lifts the suppression of the $A\otimes A$ and
$A\otimes V$ annihilation processes, while radiating a pseudoscalar
opens an $s$-wave for the $S\otimes S$ or $S\otimes P$ interaction types.

An important feature of dark ISR is that the bremsstrahlung
annihilation rate scales as $\langle\sigma v\rangle_{\rm ISR} \propto
\mathcal{O}(1/\Lambda^4)$, i.e., the same order in $\Lambda$ as the
2-body annihilation $\overline\chi\chi \rightarrow \overline{f}f$.  In
comparison, the opening of an unsuppressed $s$-wave via FSR or VIB of
a SM particle (e.g. as in the well-studied $\overline\chi\chi
\rightarrow \overline{f}f\gamma$ process) occurs only at higher order
in $1/\Lambda$, with a cross section scaling as $\langle\sigma
v\rangle_{\rm FSR, \,ISR} \propto \mathcal{O}(1/\Lambda^8)$.

When introducing a new field for dark ISR, additional competing
annihilation processes are induced.  For the ISR of a dark vector or
axial-vector, a competing $s$-wave annihilation process is
$\overline\chi\chi\rightarrow Z'Z'$.  For the case of scalar or
pseudoscalar ISR, there is no equivalent $2\rightarrow 2$ $s$-wave
process.  However, the annihilation to three pseudoscalars,
$\overline\chi\chi\rightarrow\phi\phi\phi$, is $s$-wave and can
dominate in some regions of parameter space. As such, the interplay
of several annihilation processes must always be considered.

The introduction of dark radiation should not be viewed as an
additional or unnecessary complication to dark sector theories. It is
highly likely that the dark sector, like the visible sector, has
multiple field content.  Dark vectors arise naturally when the DM
stability is due to a charge under a new gauge group, while dark
scalars are well motivated when considering mass generation in the
dark sector.  Indeed, self-consistent, gauge invariant models
frequently contain such features.  We have shown that it is important
to include dark radiative corrections in these scenarios, as they can
be the dominant annihilation channel.\\ \\

\section*{Acknowledgments}
RKL thanks the Niels Bohr International Academy for their hospitality during the completion of this work. NFB, YC and RKL were supported in part by the Australian Research Council. TJW is supported in part by the Department of Energy (DoE) Grant No. DESC-0011981. Feynman diagrams are drawn using {\sc TikZ-Feynman} \cite{Ellis:2016jkw}. \vspace{-1mm}\\

\section*{Appendix}
\appendix
\section{Three pseudoscalar cross section}
\label{sec:full3pseudo}

The cross section for DM annihilation into three pseudoscalars,
$\overline\chi\chi\rightarrow \phi\phi\phi$, is given by
\begin{align}
&\langle \sigma v\rangle_{\overline\chi\chi\rightarrow \phi\phi\phi}= 
\frac{g_\phi^6}{2^5  3!\pi^3} \int_{\rho_\phi}^\frac{1-\rho_\phi}{2} d x_2
\int_{x_1^{\rm min}}^{x_1^{\rm max}} d x_1
 \\
& \times\frac{\left( 1 -4 (x_1+x_2)+4 x_1 x_2 +4(x_1^2+x_2^2)+2\rho_\phi^2-3\rho_\phi^4\right)^2
}{8 \, m_\chi^2 \, (x_1 -\rho_\phi^2)^2\, (x_2-\rho_\phi^2)^2\, (1-x_1-x_2-\rho_\phi^2)^2} ,\nonumber 
\label{eq:full3pseudo}
\end{align}
with $x_{1,2}=E_{1,2}/2m_\chi$ where $E_{1,2}$ is the energy of two of the three pseudoscalars in the center of mass frame of the DM pair. 
The integrand is clearly symmetric with respect to $x_1$ and $x_2$, as expected.
The integration limits $x_1^{\rm min,max}$ are functions of $x_2$ and $\rho_\phi$ expressed as
\begin{equation}
x_1^{\rm min,max}=\frac{1 + 2 x_2^2+\rho_\phi^2 -x_2(3+\rho_\phi^2) \mp A}{2(1-2 x_2+\rho_\phi^2)} \;,
\end{equation}
where
\begin{align}
 A&= \Big[4 x_2^4 
 - 4 x_2(x_2^2-\rho_\phi^2)(1-\rho_\phi^2)\nonumber\\
  &+x_2^2(1-6\rho_\phi^2-3\rho_\phi^4)
  - \rho_\phi^2(1-2\rho_\phi^2-3\rho_\phi^4)\Big]^{1/2}\nonumber.
\end{align}
In the limit $m_\phi\ll m_\chi$, this produces the $s$-wave cross section shown in Eq.~(\ref{eq:massless3pseudo}).

\bibliography{darkradiation.bib}

\end{document}